%% file: paper.tex
\documentclass{article}
\usepackage{spconf}
\usepackage{mathtools}
\usepackage{amsmath,amsfonts,amssymb}
\usepackage{graphicx}
\usepackage{tikz}
\usepackage{enumitem}
\usepackage{booktabs}
\usepackage{siunitx}
\usepackage{fontawesome}
\usepackage{bbding}

\PassOptionsToPackage{hyphens}{url}
\usepackage[hyphens]{url}
\usepackage{hyperref}
\usepackage[hyphenbreaks]{breakurl}
\usepackage[capitalize]{cleveref}

\usetikzlibrary{calc,positioning,arrows,fit,backgrounds,decorations.pathreplacing}

\newcommand{\tbf}{\fontseries{b}\selectfont}

\DeclareMathOperator*{\argmin}{arg\,min}

\DeclareMathOperator*{\sg}{sg}

\DeclareSIUnit{\nothing}{\relax}
\DeclareSIUnit{\bpm}{BPM}

\crefformat{footnote}{#2\footnotemark[#1]#3}

\newlist{compactitem}{itemize}{3}
\setlist[compactitem]{topsep=0pt,partopsep=0pt,itemsep=0pt,parsep=0pt}
\setlist[compactitem,1]{label=\textbullet}
\setlist[compactitem,2]{label=---}
\setlist[compactitem,3]{label=*}

\newlist{compactdesc}{description}{3}
\setlist[compactdesc]{topsep=0pt,partopsep=0pt,itemsep=0pt,parsep=0pt}

\newlist{compactenum}{enumerate}{3}
\setlist[compactenum]{topsep=0pt,partopsep=0pt,itemsep=0pt,parsep=0pt}
\setlist[compactenum,1]{label=\arabic*}
\setlist[compactenum,2]{label=\alph*}
\setlist[compactenum,3]{label=\roman*}

\title{Self-Supervised VQ-VAE For One-Shot Music Style Transfer}

\name{Ond\v{r}ej C\'ifka$^{\star \dagger}$\thanks{This work was supported by the European Union’s Horizon 2020 research and innovation programme under the Marie Skłodowska-Curie grant agreement No.\ 765068 (MIP-Frontiers).} %
\qquad Alexey Ozerov$^{\dagger}$ \qquad Umut \c{S}im\c{s}ekli$^{\ddagger\star}$ \qquad Ga\"el Richard$^{\star}$}
\address{%
$^{\star}$LTCI, T\'el\'ecom Paris, Institut Polytechnique de Paris, %
France\\
$^{\dagger}$InterDigital R\&D, %
Cesson-S\'evign\'e,
France\quad
$^{\ddagger}$Inria/ENS, Paris, France
}

\usepackage{xcolor}
\definecolor{darkgreen}{rgb}{0,0.5,0}
\definecolor{darkred}{rgb}{0.55,0,0}
\usepackage[normalem]{ulem}

\def\preprint{}

\ifdefined\preprint
\fi

\begin{document}
\ninept
\maketitle
\begin{abstract}
Neural style transfer, allowing to apply the artistic style of one image to another, has become one of the most widely showcased computer vision applications shortly after its introduction.
In contrast, related tasks in the music audio domain remained, until recently, largely untackled.
While several style conversion methods tailored to musical signals have been proposed, most lack the `one-shot' capability of classical image style transfer algorithms.
On the other hand, the results of existing one-shot audio style transfer methods on musical inputs are not as compelling.
In this work, we are specifically interested in the problem of \emph{one-shot timbre transfer}.
We present a novel method for this task, based on an extension of the vector-quantized variational autoencoder (VQ-VAE), along with a simple self-supervised learning strategy designed to obtain disentangled representations of timbre and pitch.
We evaluate the method using a set of objective metrics and show that it is able to outperform selected baselines.
\end{abstract}

\section{Introduction}
\ifdefined\preprint%
    \begin{tikzpicture}[overlay, remember picture]
    \node at (current page.north) [yshift=-1.5cm,text width=\textwidth] {\footnotesize \textcopyright{} 2021 IEEE. Personal use of this material is permitted. Permission from IEEE must be obtained for all other uses, in any current or future media, including reprinting/republishing this material for advertising or promotional purposes, creating new collective works, for resale or redistribution to servers or lists, or reuse of any copyrighted component of this work in other works.\quad%
    DOI: \href{https://doi.org/10.1109/ICASSP39728.2021.9414235}{10.1109/ICASSP39728.2021.9414235}};
    \end{tikzpicture}%
\fi%
Neural style transfer techniques, originally proposed for images \cite{Gatys2016ImageNetworks,huang2017arbitrary},
allow applying the `artistic style' of one image to another.
Recently, there has been increased interest in developing similar methods for music, and promising works in this domain have begun to emerge.
Especially compelling are results achieved by several recent works on timbre conversion \cite{Mor2018,Huang2018TimbreTronAW,engel2020ddsp,bitton2020vector}, leading to entertaining applications.\footnote{\url{https://g.co/tonetransfer}}
However, a common property of these deep learning-based methods is that they require training for each individual target instrument.
Consequently, the set of target instruments available in these systems is typically small, as adding new ones is a time-consuming process which depends on the availability of clean training data.

In the present work, we instead propose to tackle a more general task, which we refer to as \emph{one-shot timbre transfer}.\footnote{Similarly to \cite{Cifka2020Groove2Groove}, we supplement the somewhat ambiguous term `timbre transfer' with the attribute `one-shot' to specify that we aim to imitate the timbre of \emph{one single example} presented at test time.}
Borrowing the terminology of image style transfer, our goal is to transfer the timbre of a \emph{style input} onto a \emph{content input} while preserving the pitch content of the latter.
To this end, we develop a single generic model capable of encoding pitch and timbre separately and then combining their representations to produce the desired output.

Unlike many previous music style transformation works (e.g.\ \cite{Cifka2020Groove2Groove,engel2020ddsp,wang2020learning,nercessian2020zeroshot}), we neither assume the training data to be paired or otherwise annotated, nor do we rely on existing models or algorithms to create artificial annotations (e.g.\ pitch contours or timbre-related descriptors).
This leads to the need for data-driven \emph{disentanglement} of the pitch and timbre representations learned by the model.
In this work, we propose to perform this disentanglement using a combination of discrete representation learning (via an extension of the vector-quantized variational autoencoder, or VQ-VAE \cite{Oord2017NeuralLearning}), self-supervised learning, and data augmentation.

Our contributions can be summarized as follows:
\begin{compactitem}
    \item We present the first neural model for one-shot instrument timbre transfer. The model operates via mutually disentangled pitch and timbre representations, learned in a self-supervised manner without the need for annotations.
    \item We train and test our model on a dataset where each recording contains a single, possibly polyphonic instrument.
    Using a set of newly proposed objective metrics, we show that the method constitutes a viable solution to the task, and is able to compete with baselines from the literature.
    We also provide audio examples for perceptual comparison by the reader.\footnote{\url{https://adasp.telecom-paris.fr/s/ss-vq-vae}\label{foot:website}}
    \item Since our approach to disentanglement is largely data-driven, it should be extensible to other music transformation tasks, such as arrangement or composition style transfer.
    \item Our source code is available online.\footnote{\url{https://github.com/cifkao/ss-vq-vae}}
\end{compactitem}

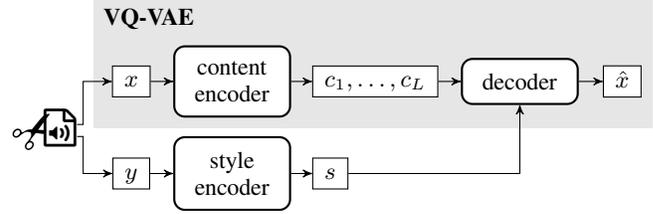
\begin{figure}
    \centering%
    \input{figs/method.tex}%
    \caption{%
        A high-level depiction of the proposed method.
        We extract pairs of segments from audio files and use them for
        self-supervised learning of a VQ-VAE with an additional style encoder.
        The content representation $c_1,\ldots,c_L$ is discrete, the style
        representation $s$ is continuous.
    }
    \label{fig:method}
\end{figure}

\section{Related Work}
\label{sec:related-work}

Prior work on our topic is rather limited.
To our knowledge, most existing works that fall under our definition of one-shot music timbre transfer \cite{Driedger2015LetIB,Tralie2018CoverSS,Foroughmand2018MusicRU} are based on non-negative matrix factorization (NMF) combined with \emph{musaicing} \cite{Zils2001MUSICALM} (a form of concatenative synthesis).
Other works on audio style transfer \cite{ulyanov2016,Grinstein2018AudioST} adapt the original image style transfer algorithm \cite{Gatys2016ImageNetworks} to audio, but focus on `sound textures' rather than timbre. In particular, the style representation used in these works is heavily pitch-dependent, which makes it unsuitable for representing musical timbre.
Finally, \cite{nercessian2020zeroshot} focuses on singing voice conversion, adopting vocoder-based techniques developed for speech. 

Speaking more generally of \emph{timbre} or \emph{style conversion}, several methods were recently proposed for musical audio \cite{Mor2018,Huang2018TimbreTronAW,engel2020ddsp,bitton2020vector}.
While these approaches achieve remarkable output quality,
they cannot be considered one-shot as
they only allow for conversion to the (small) set of styles present in the training data.
Moreover, unlike our methods, they require training a separate decoder for each target style;
in particular, \cite{Mor2018} report unsuccessful attempts to train a single decoder conditioned on the identity of the target instrument.

Other recent works \cite{philippe_esling_2018_1492373,yin_jyun_luo_2019_3527918,bitton2019assisted,luo2020unsupervised} are related to ours in that they also learn a continuous timbre representation which allows for audio generation, but are limited to the simple case of isolated notes.

\section{Background}
\label{sec:background}
\subsection{Vector-quantized autoencoder (VQ-VAE)}
\label{sec:bg-vqvae}
The VQ-VAE \cite{Oord2017NeuralLearning} is an autoencoder with a discrete latent representation.
It consists of an encoder, which maps the input $x$ to a sequence $z$ of discrete codes from a codebook, and a decoder, which tries to map $z$ back to $x$.
Using discrete latent codes places a limit on the amount of information that they can encode. %
The authors successfully exploit this property to achieve voice conversion. %
In this work, we follow a similar path to achieve music style transfer.

Formally, the encoder first outputs a sequence $E(x)\in\mathbb{R}^{L\times D}$
of $D$-dimensional feature vectors, which are then passed through a quantization (discretization) operation $Q$
which selects the nearest vector from a discrete embedding space (codebook) $e\in\mathbb{R}^{K\times D}$:
\begin{equation} 
z_i = Q(E_i(x)) = \argmin_{e_j, 1\leq j\leq K} \bigl\Vert E_i(x) - e_j\bigr\Vert.
\end{equation}

The model is trained to minimize a reconstruction error $\mathcal{L}_\text{ae}$ between the input $x$ and the output of the decoder $D(Q(E(x)))$.
The backpropagation of its gradient through the discretization bottleneck $Q$ to the encoder is enabled via \emph{straight-through estimation}, where the gradient with respect to $Q(E(x))$ received from the decoder is instead assigned to $E(x)$.
To ensure the alignment of the codebook $e$ and the encoder outputs $E(x)$, two other terms appear in the VQ-VAE objective~-- the codebook loss and the commitment loss:
\begin{align}
    \mathcal{L_\text{cbk}}&=\bigl\Vert\sg\bigl[Q(E(x))\bigr]-E(x)\bigr\Vert^2,\\
    \mathcal{L_\text{cmt}}&=\bigl\Vert Q(E(x))-\sg\bigl[E(x)\bigr]\bigr\Vert^2. 
\end{align}
Here $\sg[\cdot]$ stands for the `stop-gradient' operator, defined as identity in the forward computation, but blocking the backpropagation of gradients.
The two losses are therefore identical in value, but the first only affects (i.e.\ has non-zero partial derivatives w.r.t.)\ the codebook $e$ (via $Q$), while the second only affects the encoder $E$.
A weighting hyperparameter $\beta$ is applied to $\mathcal{L_\text{cmt}}$ in the total loss:
\begin{equation}
    \mathcal{L} = \mathcal{L}_\text{ae} + \mathcal{L}_\text{cbk} + \beta\mathcal{L}_\text{cmt}
    \label{eq:vq-vae-loss}
\end{equation}

\subsection{Self-supervised learning}
Self-supervised learning is a technique for learning representations
of unlabeled data.
The basic principle is to expose the inner structure of the data~-- by splitting each example into parts or by applying simple transformations to it~-- and then exploit this structure to define an artificial task (sometimes called the \emph{pretext task}) to which supervised learning can be applied.
Notable examples include predicting context (e.g.\ the neighboring words in a sentence \cite{mikolov2013efficient} or a missing patch in an image \cite{Pathak_2016_CVPR}), the original orientation of a rotated image \cite{gidaris2018unsupervised} or the `arrow of time' in a (possibly reversed) video \cite{Wei_2018_CVPR}.
In this work, we extract pairs of excerpts from audio files and rely on them to learn a style representation as detailed in the following section.

\section{Method}
\label{sec:method}

Given the goal of mapping two inputs~-- the content input $x$ and the style input $y$~-- to an output, it is natural to define an encoder-decoder model with two encoders (one for each input) and a single decoder.
It remains to describe how to train this model, and in particular, how to ensure the mutual disentanglement of the style and content features.
Our proposal, illustrated in \cref{fig:method}, rests on two key points:
\begin{enumerate}[label=$(\roman*)$]
    \item We use a \emph{discrete} representation $c_1,\ldots,c_L$ for content
    and train the model to reconstruct the content input, $x$;
    hence, the content encoder together with the decoder form a VQ-VAE.
    This is motivated by the success of the VQ-VAE on voice conversion
    as mentioned in \cref{sec:bg-vqvae}.
    \item The output of our style encoder is a single \emph{continuous-valued} embedding vector $s$. To ensure that the style encoder only encodes style (i.e.\ to make it content-independent), we employ a simple self-supervised learning strategy where we feed a different input $y$ to the style encoder such that $x$ and $y$ are different segments of the same audio recording (with some data augmentation applied; see \cref{sec:data-mining} for details).
\end{enumerate}
These choices are complementary to each other, as we will now see.

Firstly, $(i)$ necessarily means that the content encoder will drop some information from the content representation $c$.
Since this alone does not guarantee that only content information will be preserved, (ii) is introduced to guide the encoder to do so. Our reasoning is that
providing a separate style representation, not constrained by the discretization bottleneck, should make it unnecessary to also encode style information in $c$. 
Secondly, it can be expected that in a trained model, only information useful for reconstructing $x$ will influence the output. Hence, due to $(ii)$ and provided that $x$ and $y$ do not share any content information, we expect $s$ to only encode style.
Also note that the discretization bottleneck in $(i)$ is key for learning a useful style representation $s$: without it, $y$ may be completely ignored by the model. 

Once trained, the model is used for inference simply by feeding the content input and the style input to the respective encoders.

\subsection{Data}
\label{sec:data-mining}
Our self-supervised learning strategy consists in training on pairs of segments $x,y$ where each such pair comes from a single recording. The underlying assumption is that such $x$ and $y$ have the same style (timbre) but different content.
We combine data from two different sources, chosen to easily satisfy this assumption:
\begin{enumerate}
    \item \textbf{LMD.}\; The `full' version of the Lakh MIDI Dataset\footnote{\url{https://colinraffel.com/projects/lmd/}} \cite{raffel2016learning} (\texttt{LMD-full}), containing \SI{178}{\kilo\nothing} MIDI files (about a year's worth of music in a symbolic representation).
    We pick a random non-drum part from each file, sample two 8-second segments of this part and render them as audio using a sample-based synthesizer (FluidSynth), with the SoundFont picked randomly out of 3 options.\footnote{\emph{Fluid R3 GM}, \emph{TimGM6mb}, and \emph{Arachno SoundFont}; see \cite{soundfont-list}}

    \item \textbf{RT.}\; A set of audio tracks from PG Music;\footnote{\url{https://www.pgmusic.com}} specifically, the \num{1526} RealTracks included with Band-in-a-Box UltraPAK 2018.
    Each RealTrack (RT) is a collection of studio recordings of a single instrument playing either an accompaniment part or a solo in a single style.
    We extract pairs of short segments totalling up to \SI{20}{\minute} per RT, and clip each segment to \SI{8}{\second} after performing data augmentation (see below).
\end{enumerate}

We perform two kinds of data augmentation.
Firstly, we transpose each segment from LMD up or down by a random interval (up to 5 semitones) prior to synthesis; this ensures that the two segments in each pair have different content, but does not affect their timbre.

Secondly, we apply a set of random timbre-altering transformations to increase the diversity of the data:
\begin{itemize}
    \item \emph{(LMD only.)} Randomly changing the MIDI program (instrument) to a different one from the same broad family of instruments (keyboards \& guitars; basses; winds \& strings; \dots) prior to synthesis.
    \item \emph{(RT only.)} Audio resampling, resulting in joint time-stretch\-ing and transposition by up to $\pm4$ semitones.
    \item \num{0}--\num{4} audio effects, %
    drawn from reverb, overdrive, phaser, and tremolo, with randomly sampled parameters.
\end{itemize}
An identical set of transformations is applied to both examples in each pair to ensure that their timbres do not depart from each other.

After this procedure, we end up with \SI{209}{\kilo\nothing} training pairs
(\SI{119}{\kilo\nothing} from LMD\footnote{The final number is lower than the number of files in LMD due to corrupt MIDI files and parts with insufficiently many notes being discarded.}
and \SI{90}{\kilo\nothing} from RT).

\subsection{Model and training details}
\label{sec:model-details}
We represent the audio signal as a log-scale magnitude STFT (short-time Fourier transform) spectrogram with a hop size of $\SI{1/32}{\second}$ and \num{1025} frequency bins.
To obtain the output audio, we invert the STFT
using the Griffin--Lim algorithm \cite{griffin1984signal}.

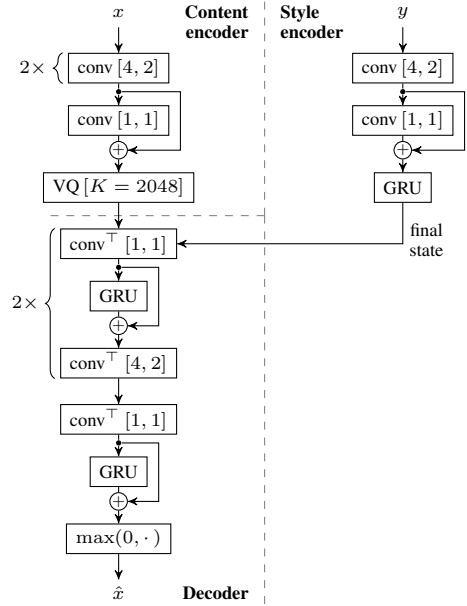
\begin{figure}
    \centering%
    \input{figs/architecture.tex}%
    \caption{%
        The model architecture. All convolutions are 1D, with the kernel size and stride shown.
        All layers have \num{1024} channels, except for the last two ($\text{conv}^\top$ \& $\text{GRU}$), which have \num{1025} (the number of frequency bins).
        All layers except for the input layers and the VQ are preceded by batch normalization and a Leaky ReLU activation \cite{maas2013rectifier}. $\text{conv}^\top$ stands for transposed convolution.
    }
    \label{fig:architecture}
\end{figure}

The model architecture is depicted in \cref{fig:architecture}.
The encoders treat the spectrogram as a 1D sequence with \num{1025} channels and process it using a series of 1D convolutional layers which serve to downsample it (i.e.\ reduce its temporal resolution).
The last layer of the style encoder is a GRU (gated recurrent unit \cite{Cho2014LearningPR}) layer, whose final state $s$ (a \num{1024}-dimensional vector) is used as the style representation.
This vector $s$ is then fed to the $1^\text{st}$ and the $4^\text{th}$ decoder layer by concatenating it with the preceding layer's outputs at each time step.

The decoder consists of 1D transposed convolutional layers which upsample the feature sequence back to the original resolution.
GRU layers are inserted %
to combine the content and style representations in a context-aware fashion.

We train the model using Adam \cite{Kingma2015AdamAM}
to minimize the VQ-VAE loss from \cref{eq:vq-vae-loss}, defining the reconstruction loss $\mathcal{L}_\text{ae}$ as the mean squared error between $x$ and $\hat x$.
We train for 32 epochs, taking about 20 hours in total on a Tesla V100 GPU.

\section{Experiments}
As in previous music style transformation work \cite{Cifka2019SupervisedSM,Cifka2020Groove2Groove},
we wish to evaluate our method on two criteria: (a) content preservation and (b) style fit.
In timbre transfer, these should express (a) how much of the pitch content of the content input is retained in the output, and (b) how well the output fits the target timbre.
To this end, we propose the following objective metrics for measuring pitch and timbre dissimilarity, respectively, between an output and a reference recording:
\begin{enumerate}[label=(\alph*)]
    \item \textbf{Pitch:} We extract pitch contours from both recordings using a multi-pitch version of the MELODIA algorithm \cite{salamon2012melody} implemented in the Essentia library \cite{bogdanov2013essentia}.
    We round the pitches to the nearest semitone and express the mismatch between the two pitch sets $A,B$ at each time step as the Jaccard distance:
    \begin{equation*}
        d_J(A,B) = 1-\frac{\lvert A\cap B\rvert}{\lvert A\cup B\rvert}
    \end{equation*}
    We report the mean value of this quantity over time.

    \item \textbf{Timbre:} Mel-frequency cepstral coefficients (MFCCs) 2--13 are generally considered to be a good approximate timbre representation \cite{GJR13}.
    Since they are computed on a per-frame basis, %
    we train a triplet network \cite{hoffer2015deep}
    on top of them to aggregate them over time and output a single dissimilarity score.
    More details can be found on the supplementary website.\cref{foot:website}
\end{enumerate}
We compare our method to 2 trivial baselines and 2 baselines from the literature:
\begin{compactitem}
    \item \textsc{cp-content}: Copies the content input to the output.
    \item \textsc{cp-style}: Copies the style input to the output.
    \item \emph{U+L}: The algorithm of Ulyanov and Lebedev \cite{ulyanov2016} (not specifically designed for timbre transfer), consisting in optimizing the output spectrogram for a content loss and a style loss.
    We tune the ratio of the weights of the two losses
    on a small synthetic validation set to minimize the log-spectral distance (LSD) to the ground truth (see \cref{sec:synth-test}).
    \item \emph{Musaicing}: A freely available implementation\footnote{\url{https://github.com/ctralie/LetItBee/}} of the musaicing algorithm of Driedger et al.\ \cite{Driedger2015LetIB}.
\end{compactitem}

\subsection{Artificial benchmark}
\label{sec:synth-test}
First, we evaluate our method on a synthetic dataset generated from MIDI files.
Although such data is not completely realistic, it enables conducting a completely objective benchmark by comparing the outputs to a synthetic ground truth.

We generate the data from the Lakh MIDI Dataset (LMD) similarly as in \cref{sec:data-mining}, but using a set of files held out from the training set, and with no data augmentation. We use the \emph{Timbres Of Heaven} SoundFont (see \cite{soundfont-list}), not used for the training set.

We randomly draw \num{721} content--style input pairs and generate a corresponding ground-truth target for each pair by synthesizing the content input using the instrument of the style input.
More details are given on the supplementary website.\cref{foot:website}

Both the pitch and timbre distance are measured with respect to the ground-truth target.
Additionally, we measure an overall distance to the target as the root-mean-square error computed on \si{\decibel}-scale mel spectrograms;
this is known as the \emph{log-spectral distance} or \textbf{LSD}.

\begin{table}
    \centering%
    \footnotesize%
    \renewcommand{\arraystretch}{0.9}%
    \begin{tabular}{lrrrrr}
    \toprule
    & \multicolumn{3}{c}{Artificial} & \multicolumn{2}{c}{Real} \\
    \cmidrule(lr){2-4}
    \cmidrule(l){5-6}
    System & \multicolumn{1}{c}{LSD\textsubscript{T}} & \multicolumn{1}{c}{Timbre\textsubscript{T}} & \multicolumn{1}{c}{Pitch\textsubscript{T}} & \multicolumn{1}{c}{Timbre\textsubscript{S}} & \multicolumn{1}{c}{Pitch\textsubscript{C}} \\
    \midrule
    \textsc{cp-content} & 14.62 & 0.3713 & 0.5365 & 0.4957 & $\frac{\hphantom{0.0000}}{}$  \\
    \textsc{cp-style}  & 20.36 & 0.2681 & 0.8729 & $\frac{\hphantom{0.0000}}{}$ & 0.9099 \\
    \cmidrule{1-6}
    U+L \cite{ulyanov2016} & 14.50 & 0.3483 & \tbf 0.5441 & 0.4792 & \tbf 0.1315 \\
    Musaicing \cite{Driedger2015LetIB} & 14.51 & 0.2933 & 0.6445 & 0.2319 & 0.6297 \\
    This work & \tbf 12.16 & \tbf 0.2063 & 0.5500 & \tbf 0.2278 & 0.6197 \\
    \bottomrule
    \end{tabular}

    \caption{Evaluation results. Distances marked S, C, and T are computed w.r.t.\ the \uline{s}tyle input, the \uline{c}ontent input, and the synthetic  \uline{t}arget, respectively.
    Results that are trivially 0 are omitted.%
    }
    \label{tab:results}
\end{table}

\subsection{`Real data' benchmark}
\label{sec:real-test}
We create a more realistic test set based on the `Mixing Secrets' audio library \cite{mixing-secrets}, containing over \num{400} multi-track recordings from various (mostly popular music) genres.
After filtering out multi-instrument, vocal and unpitched percussion tracks, we extract 690 content-style input pairs similarly as in \cref{sec:synth-test}.
As no ground truth is available in this dataset, we compute the pitch and timbre metrics with respect to the content and style input, respectively.

\subsection{Results}
The results of both benchmarks are shown in \cref{tab:results}.
First, our system outperforms all baselines on LSD and the timbre metric. %
The difference to the \textsc{cp-content} baseline is negative in more than \SI{75}{\percent} of examples on both of these metrics and in both benchmarks.
Hence, viewing our system as a timbre transformation applied to the content input, we can conclude that, informally speaking, the transformation is at least partly successful in more than \SI{75}{\percent} of cases.
We may also notice that the result of \textsc{cp-style} on timbre is, somewhat counter-intuitively, outperformed by our system.
This may be a sign that the timbre metric is still somewhat influenced by pitch. %

Turning to the pitch distance metric, we note that its values seem rather high ($>\num{0.5}$ on a scale from 0 to 1).
However, most of this error should be attributed to the pitch tracking algorithm rather than to the systems themselves.
This is documented by the fact that the pitch distance of \textsc{cp-content} to the ground-truth target is \num{0.54} instead of \num{0}.
Another useful value to look at is the result of \textsc{cp-style}: as the style input is selected randomly, its pitch distance value should be high, and is indeed close to \num{0.9}.
Using these two points of reference, we observe that our system's result is much closer to the former than to the latter in both benchmarks, which is the desired outcome.
Moreover, it outperforms the musaicing baseline in both cases, albeit only slightly on real inputs.

\section{Discussion}

Our subjective observations upon examining the outputs\cref{foot:website} mostly match the objective evaluation. We find that, although the sound quality of our outputs is not nearly perfect, their timbre typically does sound much closer to the style input than to the content input.
(Low synthesis quality and various artifacts are somewhat expected, as they are a common occurrence with the Griffin-Lim algorithm, as well as decoders based on transposed convolutions \cite{pons2020upsampling}. However, synthesis quality is not the main focus of this preliminary work.)

The pitch of the content input is generally well preserved in the output, yet faster notes and polyphony seem to pose a problem.
We believe this is caused by a low capacity of the discrete content representation.
Even though a codebook size of \num{2048} seems more than sufficient in theory, we found that on both of our test sets combined, only \num{81} of the codebook vectors are actually used in practice. This means, for example, that at a tempo of \SI{120}{\bpm}, only \num{25.4} bits of information can be encoded per beat.
This `codebook collapse' \cite{Dieleman2018TheScale} is a known issue with VQ-VAEs.

We also observe that our method works better on target instruments with a temporally `stable' sound, e.g.\ piano;
this might also explain why our method achieves better evaluation results on synthetic inputs (generated using samples) than on real ones, which are less predictable.
A likely culprit is our use of a deterministic model, which cannot possibly capture the acoustic variability of instruments like saxophone or violin while being able to convert from an instrument that lacks this variability.
This could be remedied by replacing our decoder with a probabilistic one which models a fully expressive conditional distribution, such as WaveNet \cite{van2016wavenet}.

The musaicing baseline, which uses fragments from the style input to construct the output, generally matches the target timbre very precisely, but is often less musically correct than ours.
For example, note onsets tend to lack clear attacks; pitch errors and spurious notes occur, especially when the style input is non-monophonic or fast.

Finally, let us comment on the U+L baseline.
Although its results on pitch are excellent, this is caused by the fact that the style weight obtained by tuning is very low (about \num{100} times lower than the content weight), causing the algorithm to behave much like \textsc{cp-content}.
This is also reflected by the timbre metric.
Experimenting with higher weights, we notice that the algorithm is able to transfer fragments of the style input to the output, but cannot transpose (pitch-shift) them to match the content input.

\section{Conclusion}
We have proposed a novel approach to one-shot timbre transfer, based on an extension of the VQ-VAE, along with a simple self-supervised learning strategy.
Our results demonstrate that the method constitutes a viable approach to the timbre transfer task and is able to outperform baselines from the literature.

The most important shortcoming of our method seems to be the use of a deterministic decoder.
We believe that a more expressive decoder such as WaveNet should allow improving the performance especially on instruments with great temporal variability, and perhaps enable extensions to more challenging style transfer tasks, such as arrangement or composition style transfer.

\bibliographystyle{IEEEbib}
\bibliography{refs}

\end{document}

%% file: figs/method.tex
{
\small
\begin{tikzpicture}[x=0.4cm,y=0.4cm,auto]
    \tikzset{
        >=stealth',
        module/.style={
           rectangle,
           rounded corners,
           draw=black, thick,
           fill=white,
           text width=4.2em,
           inner ysep=0.6em,
           text centered},
        data/.style={
           rectangle,
           draw=black,
           fill=white,
           inner xsep=0.5em,
           inner ysep=0.2em,
           text centered},
    }

    \node[data] (cinput) {$x$\vphantom{dp}};
    \node[module,align=center,right=0.75 of cinput] (cenc) {content\\encoder}
        edge[<-] (cinput);
    \node[data,right=0.7 of cenc] (crep) {$c_1,\ldots,c_L$\vphantom{dp}}
        edge[<-] (cenc);

    \node[data,below=2 of cinput] (sinput) {$y$\vphantom{dp}};
    \node[module,align=center] at (cenc |- sinput) (senc) {style\\encoder}
        edge[<-] (sinput);
    \node[data,right=0.7 of senc] (srep) {$s$\vphantom{dp}}
        edge[<-] (senc);

    \node[module,right=0.75 of crep] (dec) {decoder}
        edge[<-] (crep);
    \draw[->] (srep) -| (dec);

    \node[data,right=0.8 of dec] (out) {$\hat x$\vphantom{dp}}
        edge[<-] (dec);
    
    \begin{scope}[on background layer]    
        \node[fill=black!10,
              fit={
                ($(cinput.west)+(-0.25,0)$) (out)
                (cenc) ($(cenc.north)+(0,1.5em)$)
              },
              inner sep=4]
            (vqvae) {};
        \node[below right=0.08 of vqvae.north west] () {\bf VQ-VAE};
    \end{scope}
    
    \node[left=1.8 of $(cinput)!0.5!(sinput)$, inner sep=0] (audiofile) {\Large\rotatebox[origin=b]{45}{\ScissorRightBrokenBottom}\hskip-0.48em\faFileAudioO};
    \draw[->] ($(audiofile.east)+(0,0.1)$) -- ($(audiofile.east)+(0.2,0.1)$) |- (cinput.west);
    \draw[->] ($(audiofile.east)+(0,-0.3)$) -- ($(audiofile.east)+(0.2,-0.3)$) |- (sinput.west);
\end{tikzpicture}%
}%

%% file: figs/architecture.tex
{%
\scriptsize
\begin{tikzpicture}[x=0.5cm,y=0.17cm,auto]
    \tikzset{
        >=stealth',
        layer/.style={
           rectangle,
           draw=black,
           fill=white,
           inner xsep=0.5em,
           inner ysep=0,
           minimum height=1.6em,
           text centered}
    }

    \node[minimum height=1.5em] (x) {$x$};
    \node[layer,below=2 of x] (ce1) {$\text{conv}\,[4,2]$} edge[<-] (x);
    \draw [decorate,decoration={brace,amplitude=0.5em},transform canvas={xshift=-2}]
        (ce1.south west) -- (ce1.north west)
        node [midway,xshift=-0.5em] {$2\times$};
    
    \node[circle,fill,inner sep=0.1em,below=0.4 of ce1] (ce3a) {} edge (ce1);
    \node[layer,below=1.7 of ce1] (ce3) {$\text{conv}\,[1,1]$} edge[<-] (ce3a);
    \node[circle,draw,fill=white,below=0.4 of ce3,inner sep=0] (ce3b) {\tiny$+$} edge (ce3);
    \draw[->] (ce3a) -- ([xshift=0.6em]ce3a -| ce3.east) |- (ce3b);

    \node[layer,below=1 of ce3b] (ce4) {$\text{VQ}\,[K=2048]$} edge[<-] (ce3b);

    \node[minimum height=1.5em,right=14em of x] (y) {$y$};
    \node[layer] at (y |- ce1) (se1) {$\text{conv}\,[4,2]$} edge[<-] (y);

    \node[circle,fill,inner sep=0.1em,below=0.4 of se1] (se2a) {} edge (se1);
    \node[layer,below=1.7 of se1] (se2) {$\text{conv}\,[1,1]$} edge[<-] (se2a);
    \node[circle,draw,fill=white,below=0.4 of se2,inner sep=0] (se2b) {\tiny$+$} edge (se2);
    \draw[->] (se2a) -- ([xshift=0.6em]se2a -| se2.east) |- (se2b);

    \node[layer,below=1 of se2b] (se3) {$\text{GRU}$} edge[<-] (se2b);

    \node[layer,below=2 of ce4] (d1) {$\text{conv}^\top\,[1,1]$};  
    \draw[->] (ce4) -- (d1);
    \draw[->] (se3) |- node [right,pos=0.45,align=center] {final\\state} (d1);

    \node[circle,fill,inner sep=0.1em,below=0.4 of d1] (d2a) {} edge (d1);
    \node[layer,below=1.7 of d1] (d2) {$\text{GRU}$} edge[<-] (d2a);
    \node[circle,draw,fill=white,below=0.4 of d2,inner sep=0] (d2b) {\tiny$+$} edge (d2);
    \draw[->] (d2a) -- ([xshift=0.6em]d2a -| d2.east) |- (d2b);
    
    \node[layer,below=1 of d2b] (d3) {$\text{conv}^\top\,[4,2]$} edge[<-] (d2b);
    
    \draw [decorate,decoration={brace,amplitude=0.5em},transform canvas={xshift=-2}]
        (d3.south west) -- (d1.north -| d3.west)
        node [midway,xshift=-0.5em] {$2\times$};
    
    \node[layer,below=2 of d3] (d3-1) {$\text{conv}^\top\,[1,1]$} edge[<-] (d3);
    
    \node[circle,fill,inner sep=0.1em,below=0.4 of d3-1] (d7a) {} edge (d3-1);
    \node[layer,below=1.7 of d3-1] (d7) {$\text{GRU}$} edge[<-] (d7a);
    \node[circle,draw,fill=white,below=0.4 of d7,inner sep=0] (d7b) {\tiny$+$} edge (d7);
    \draw[->] (d7a) -- ([xshift=0.6em]d7a -| d7.east) |- (d7b);
    
    \node[layer,below=1 of d7b] (d8) {$\max(0,\cdot\,)$} edge[<-] (d7b);
    
    \node[below=2 of d8] (xhat) {$\hat x$} edge[<-] (d8);

    \coordinate (center-y) at ($(ce4.south)!0.5!(d1.north)$);
    \coordinate (center-x) at ($(d1.east)!0.5!(se2.west)$);
    \coordinate (center) at (center-x |- center-y);
    \begin{scope}[on background layer] 
        \draw[dashed,gray] (center) -- (center -| ce4.west);
        \draw[dashed,gray] (center) -- (center |- x.north);
        \draw[dashed,gray] (center) -- (center |- xhat.south);
    \end{scope}
    \node[anchor=north east,align=right] at ([xshift=-0.5em]x.north -| center) () {\bf Content\\\bf encoder};
    \node[anchor=north west,align=left] at ([xshift=0.5em]x.north -| center) () {\bf Style\\\bf encoder};
    \node[anchor=south east,align=right] at ([xshift=-0.5em]xhat.south -| center) () {\bf Decoder};
\end{tikzpicture}%
}%